\documentstyle[12pt,aps]{revtex}
\begin{document}
\draft
\title{The General Decomposition Theory of $SU(2)$ Gauge Potential , Topological
Structure and Bifurcation of $SU(2)$ Chern Density }
\author{Yishi Duan\thanks{%
Email: ysduan@lzu.edu.cn }, Libin Fu\thanks{%
Corresponding author. E-mail: itp2@lzu.edu.cn.}}
\address{Institute of Theoretical Physics,Lanzhou University,Lanzhou, Gansu,\\
730000, P.R.China}
\date{\today}
\maketitle

\begin{abstract}
\begin{center}
{\bf Abstract}
\end{center}

By means of the geometric algebra the general decomposition of $SU(2)$ gauge
potential on the sphere bundle of a compact and oriented 4-dimensional
manifold is given. Using this decomposition theory the $SU(2)$ Chern density
has been studied in detail. It shows that the $SU(2)$ Chern density can be
expressed in terms of the $\delta -$function $\delta \left( \phi \right) $.
And one can find that the zero points of the vector fields $\phi$ are
essential to the topological properties of a manifold. It is shown that
there exists the crucial case of branch process at the zero points. Based on
the implicit function theorem and the taylor expansion, the bifurcation of
the Chern density is detailed in the neighborhoods of the bifurcation points
of $\phi$. It is pointed out that, since the Chren density is a topological
invariant, the sum topological chargers of the branches will remain constant
during the bifurcation process.
\end{abstract}



\section*{I. Introduction}

The topological properties of physics system play important roles in
studying some physical problem. It is well known that the gauge
potential(connection) and gauge field(curvature) is essential to establish
direct relationship between differential geometry and topological
invariants. The decomposition theory of gauge potential provides a powerful
method in researching some topological properties. It has been effectively
used to study the topological gauge theory of dislocation and disclinations
in condensed matter physics$^1$, the geometrization of Planck constant in
terms of the space time defect in General Relativity$^{2,3}$, the space-time
dislocations in the early universe$^4$, and the Gauss-Bonnet-Chern theorem$%
^{3,5}$. The essential feature of the decomposition shows that the gauge
potential have inner structure$^{6,1}$. Generally, the topological
characteristics of a manifold are represented by the properties of a smooth
vector field on it , or in other words, the smooth vector fields carry the
topological information of a manifold, which inspire us to study the
decomposition theory of the gauge potential in terms of the unit vector
field on the manifold.

It is well known that the $SU(2)$ gauge theory and the second Chern class
has been widely used in discussing many physical problem. Such as the
magnetic monopole$^{6-9}$, the anomaly in nonlinear $\sigma $-models$^{10}$.
Especially the problems on four-manifold$^{11,12}$, the instantons$^{13-15}$%
, the merons$^{16,17}$, the Donsdson theory$^{18}$, and so on. It urges us
to study the decomposition theory of the $SU(2)$ gauge potential and the
topological properties on four-manifold.

In this paper, we will establish a general decomposition theory of the $%
SU(2) $ gauge potential in terms of the unit vector $\vec n$ on the compact
and oriented 4-dimensional manifold. And by means of geometric algebra, we
can describe the unit vector $\vec n$ as an element of $Spin(3)^{19}$. One
can show that the general decomposition formula of $SU(2)$ gauge potential
has a global property. The Chern density of the principal $P(\pi ,M,SU(2))$
will be studied by using the decomposition formula. One shows that the $%
SU(2) $ Chern density takes the form of the $\delta -$function $\delta
\left( \vec \phi \right) .$ The topological structure of the $SU(2)$ Chern
density can be labeled by the Brouwer degrees and the Hopf index. And the
further research shows that there exist the crucial cases of branch process
in the topological density, when the Jacobian $D(\frac \phi x)=0.$ We
calculate out the different branches of the Chern density by using the
implicit function theorem. It is pointed out that the topological charges
will be splitted at the critical points.

This paper is arranged in five sections. In section 2 we will study a
general decomposition theory of $SU(2)$ gauge potential on a sphere bundle.
In section 3 we will study the topological structure of the $SU(2)$ Chern
density by using the decomposition expression given in section 2. The
bifurcation of the Chern density will be studied in section 4. Then there
will be a conclusion at last.

\section*{II. The\- decomposition theory of ${\it SU(2)}$ gauge potential}

In this section we will give the decomposition theory of $SU(2)$ gauge
potential in terms of the sphere bundle on a compact and oriented
4-dimensional manifold. Firstly we have to give the basic notions which are
necessary for our discussions.

Let $V$ be a unit $SU(2)$ Clifford vector

\begin{equation}
\label{1}V=V^a\sigma _a\ ,\qquad a=1,2,3\ ; 
\end{equation}
and 
\begin{equation}
\label{2}V^aV^a=1. 
\end{equation}
In which $\sigma _a$ are Pauli matrixes, and the base of $SU(2)$ Clifford
algebra$^{20}$.

The covariant derivative 1-form of $V$ is given by 
\begin{equation}
\label{3}DV=dV-[A,V] 
\end{equation}
where $A$ is the $SU(2)$ gauge potential 1-form: 
\begin{equation}
\label{4}A=i/2A^a\sigma _a\ , 
\end{equation}
and 
\begin{equation}
\label{5}A^a=A_\mu ^adx^\mu \qquad \mu =0,1,2,3. 
\end{equation}
In gauge theory, the potential 1-form undergoes the gauge transformation: 
\begin{equation}
\label{gtr}A^{\prime }=SAS^{-1}+dSS^{-1}. 
\end{equation}

In our viewpoint$^6$, the gauge potential 1-form $A$ can be decomposed and
has inner structure. The main feature of the decomposition theory of the
gauge potential is that the gauge potential $A$ can be generally decomposed
as 
\begin{equation}
\label{Dt}A=a+b, 
\end{equation}
where $a$ and $b$ are required to satisfy the gauge transformation and
vector covariant transformation rules, i.e.,$^{19}$ 
\begin{equation}
\label{a}a^{\prime }=SaS^{-1}+dSS^{-1}, 
\end{equation}
and 
\begin{equation}
\label{b}b=SbS^{-1}. 
\end{equation}
From (\ref{a}) and (\ref{b}), one can show that the gauge potential $A$
rigorously satisfies the gauge transformation%
$$
A^{\prime }=a^{\prime }+b^{\prime }=S(a+b)S^{-1}+dSS^{-1} 
$$

Let $V_{(i)}$ $(i=1,2,3)$ be an orthonormal basis of $SU(2)$ Clifford vector
with the orthogonal relations: 
\begin{equation}
\label{orthon}V_{(i)}\cdot V_{(j)}=\delta _{ij} 
\end{equation}
i.e.%
$$
V_{(i)}^aV_{(j)}^a=\delta _{ij}\ , 
$$
where $V_{(i)}\cdot V_{(j)}$ is the Clifford scalar product defined by$^{19}$
\begin{equation}
\label{scalar}V_{(i)}\cdot V_{(j)}=\frac 12(V_{(i)}V_{(j)}+V_{(j)}V_{(i)}). 
\end{equation}
Since the potential $A$ is also a $SU(2)$ Clifford vector, we have the
projection formula \qquad 
\begin{equation}
\label{prf}A=(A\cdot V_{(i)})V_{(i)}. 
\end{equation}
Substituting this formula into (\ref{3}), we obtain 
\begin{equation}
\label{dv}DV_{(i)}=dV_{(i)}-(A\cdot V_{(j)})[V_{(j)},V_{(i)}], 
\end{equation}
Using (\ref{orthon}), (\ref{scalar}) and (\ref{dv}), and considering that 
\begin{equation}
[V_{(j)},V_{(i)}]=2V_{(j)}V_{(i)}-2V_{(i)}\cdot V_{(j)}, 
\end{equation}
we can rewrite $A$ as 
\begin{equation}
\label{decomposition}A=1/4dV_{(i)}V_{(i)}-1/4DV_{(i)}V_{(i)}. 
\end{equation}
According to (\ref{Dt}), we define that 
\begin{equation}
\label{a2}a=\frac 14dV_{(i)}V_{(i)}, 
\end{equation}
and 
\begin{equation}
\label{b2}b=\frac 14DV_{(i)}V_{(i)}. 
\end{equation}
It is easy to prove that the decomposition formula above satisfies the
requirement in(\ref{gtr}), even has global property and is independent from
the local coordinates.

Let a family $\{W,V,U,\cdot \cdot \cdot \}$ be an open cover of $M$ and $%
S_{uv}$ be the transition matrix function which satisfy the following
condition$^{21}$ 
\begin{equation}
S_{uu=1},\quad S_{vu}^{-1}=S_{uv},\quad S_{wv}S_{vu}S_{uw}=I;\quad W\cap
V\cap U\neq \emptyset . 
\end{equation}
For any two open neighborhoods $V$ and $U$ , if $V\cap U\neq \emptyset ,$
then 
\begin{equation}
\label{vt}V_{(i)v}=S_{vu}V_{(i)}S_{vu}^{-1}, 
\end{equation}
where the subscripts $``_u"$ and ``$_v"$ are represent the open covers $U$
and $V$ correspondingly, and we know that the gauge potential $A$ undergoes
the gauge transformation 
\begin{equation}
\label{gr}A_v=S_{vu}A_uS_{vu}^{-1}+dS_{vu}S_{vu}^{-1}\ , 
\end{equation}
which is the fundamental condition for the existence of the gauge potential
on the principal $P(\pi ,M,SU(2)).$ In the physics terminology $S_{vu}$ is
just the gauge transformation(\ref{gtr}) in the gauge theory. In the
following , for abbreviation , we shall use the notation $S_{vu}=S$ .
According to (\ref{vt}), we deduce that 
\begin{equation}
\label{cced}%
dV_{(i)v}V_{(i)v}=SdV_{(i)u}V_{(i)u}S^{-1}+dSS^{-1}-V_{(i)v}dSS^{-1}V_{(i)v}. 
\end{equation}
Considering that $A$ is a vector of Clifford algebra,we can see that $%
dSS^{-1}$ is a vector of Clifford algebra also, then there exit the
following formula$^{19}$ 
\begin{equation}
\label{cce}V_{(i)v}dSS^{-1}V_{(i)v}=-3dSS^{-1}. 
\end{equation}
Noticing that 
$$
DV_{(i)v}=SDV_{(i)u}S^{-1}, 
$$
from (\ref{decomposition}), and making use of (\ref{vt}), (\ref{cced}) and (%
\ref{cce}), one can easily obtain 
\begin{equation}
A_v-\frac 14(dV_{(i)v}V_{(i)v}-DV_{(i)v}V_{(i)v})=S[A_u-\frac 14%
(DV_{(i)u}V_{(i)u}-DV_{(i)u}V_{(i)u})]S^{-1}. 
\end{equation}
The expression above shows that if the decomposition formula on the open
neighborhood $U$ 
$$
A_v=\frac 14(dV_{(i)v}V_{(i)v}-DV_{(i)v}V_{(i)v}) 
$$
holds true, the decomposition formula on the open neighborhood $U$ 
$$
A_u=\frac 14(dV_{(i)u}V_{(i)u}-DV_{(i)u}V_{(i)u}) 
$$
must holds true, too. This means that the general decomposition formula (\ref
{decomposition}) has a global property and is independent from the choice of
the local coordinates.

We know that the characteristic class is the fundmental topologial property,
and it is independent of the gauge potential$^{22}$. So, to disscus the
Chern class, we can take $A$ as 
\begin{equation}
\label{cd}A=\frac 14dU_{(i)}U_{(i)}. 
\end{equation}
One can regard it as a special gauge. By use of this formula, the magnetic
monopole can be studied as in$^{6-8}$. In this paper we want to discuss the
topological properties of a 4-dimensional manifold, a decomposition formula
in terms of the sphere bundle over the manifold will be convenient.

Now, let us return to discuss the decomposition theory on the sphere bundle
of a compact and oriented 4-dimensional manifold. Let $M$ be a compact and
oriented 4-dimensional manifold and $P(\pi ,M,G)$ be a principal bundle with
the structure group $G=SU(2).$ A smooth vector field $\phi ^A(A=0,1,2,3)$
can be found on the base manifold $M$. We define a unit vector field on $M$
as 
\begin{equation}
\label{nm}n^A=\phi ^A/||\phi ||\qquad A=0,1,2,3\ ; 
\end{equation}
$$
||\phi ||=\sqrt{\phi ^A\phi ^A}, 
$$
in which the superscripts ``$A$'' are the local orthonormal frame index.

In fact $\vec n$ is identified as a section of the sphere bundle over $M$
(or a partial section of the vector bundle over $M$)$^5$. We see that the
zeros of $\vec \phi $ are just the singular points of $\vec n$ . Since the
global property of a manifold has close relation with zeros of a smooth
vector fields on it, this expression of the unit vector $\vec n$ is a very
powerful tool in the discussion of the global topology. It naturally
guarantee the constraint 
\begin{equation}
\label{na}n^An^A=1.
\end{equation}
We can express the unit vector $\vec n$ in terms of Clifford algebra$^{23}$
as 
\begin{equation}
\label{n}n=n^As_A,\qquad A=0,1,2,3.
\end{equation}
where 
\begin{equation}
\label{ss}s=(I,i\vec \sigma ),\qquad s^{\dagger }=(I,-i\vec \sigma ).
\end{equation}

We know that $n$ is just an element of $Spin(3)$ in terms of geometric
algebra$^{19}$. And we can rewrite (\ref{na}) as 
\begin{equation}
\label{nomal2}nn^{\dagger }=1. 
\end{equation}

It is easy to see that $n$ has three independent components. Since the
orthonormal basis of $SU(2)$ Clifford one-vector just have three independent
components, we can express $U_{(i)}$ in terms of an element of $Spin(3)$ .
From gauge transformations of the orthonormal basis given in $^{24}$, and
the spinorial transformation given in $^{25}$, one can expand $U_{(i)}$ in
terms of an element of $Spin(3)$ as: 
\begin{equation}
\label{ex}U_{(i)}^a=2(n^0)^2\delta ^{ia}+2n^0n^b\in ^{iba}+2n^in^a-\delta
^{ia}.
\end{equation}
It is easy to prove 
\begin{equation}
U_{(i)}^aU_{(j)}^a=\delta _{ij.}
\end{equation}
Using (\ref{ex}) and (\ref{cd}), as a result, the gauge potential $A$ will
be expressed as 
\begin{equation}
\label{result}A=dnn^{\dagger }\ .
\end{equation}
Because $n$ is an unit element of the sphere bundle over $M$ , we will
discuss the topological property of the principle $P(\pi ,M,SU(2))$ directly
by using this formula.

\section*{III. The $SU(2)$ Chern Density And Its Inner Structure}

It is well known that gauge potential(connection) and gauge field(curvature)
play essential roles in discussing the topological properties of a manifold.
For the principal $P(\pi ,M,SU(2))$ the $SU(2)$ Chern density is a important
topological characteristic. In this section we will discuss the $SU(2)$
Chern density by using the decomposition formula (\ref{result}) which we
have just given in the last section.

We know that the second Chern class is the fundamental characteristic of the
principal $P(\pi ,M,SU(2))$ , it is denoted as: 
\begin{equation}
\label{ch2}C_2(P)=\frac 1{8\pi ^2}Tr(F\wedge F). 
\end{equation}
The curvature $F$ is defined as 
\begin{equation}
\label{f}F=dA-A\wedge A. 
\end{equation}

One can show that the gauge field $F$ is generalized function when there
exit $m$ singular points $z_i$ $(i=1,2,\cdot \cdot \cdot ,m)$ in the unit
vector field $n.$ Let us substitute the formula (\ref{result}) into (\ref{f}%
), then

\begin{equation}
F\left\{ 
\begin{array}{lll}
=0\qquad when & x\neq z_i &  \\ 
\neq 0\qquad when & x=z_i &  
\end{array}
\right. . 
\end{equation}
where $z_i$ are the singular points of $n$. This means that the gauge field $%
F$ vanishes at the region where $n$ has no singular points, but at the
singular points of $n$ the gauge field $F$ does not vanish. This feature we
will show soon is due to the non-triviality property of the principal $P(\pi
,M,SU(2))$ , and is essential to study some physical problem.

The second $SU(2)$ Chern class can be rewrite in terms of Chern-Simon$%
^{22,26}$ 
\begin{equation}
\label{ch2a}C_2(P)=\frac 1{8\pi ^2}d\Omega ,
\end{equation}
and 
\begin{equation}
\Omega =\frac 1{8\pi ^2}Tr\left( A\wedge dA-\frac 23A\wedge A\wedge A\right) 
\end{equation}
which is known as Chern-Simon form$^{27}$.

Substituting (\ref{result}) into (\ref{ch2a}) and considering (\ref{nomal2}%
), we obtain 
\begin{equation}
C_2(P)=\frac 1{24\pi ^2}Tr(dn\wedge dn^{\dagger }\wedge dn\wedge dn^{\dagger
}). 
\end{equation}
In detail, substituting (\ref{n}) into the formula above 
$$
C_2(P)=\frac 1{24\pi ^2}\in ^{\mu \nu \lambda \rho }\partial _\mu
n^A\partial _\nu n^B\partial _\lambda n^C\partial \rho n^DTr\left(
s_As_B^{\dagger }s_Cs_D^{\dagger }\right) dx^4 
$$
\begin{equation}
\label{tmrs}=\frac 1{12\pi ^2}\in ^{\mu \nu \lambda \rho }\in
_{ABCD}\partial _\mu n^A\partial _\nu n^B\partial _\lambda n^C\partial _\rho
n^Ddx^4. 
\end{equation}
By substituting (\ref{nm}) into (\ref{tmrs}), considering 
\begin{equation}
dn^A=\frac{d\phi ^A}{||\phi ||}+\phi ^Ad\left( \frac 1{||\phi ||}\right) , 
\end{equation}
we have 
\begin{equation}
C_2(P)=-\frac 1{4\pi ^2}\frac{\partial ^2}{\partial \phi ^A\partial \phi ^A}%
\left( \frac 1{||\phi ||^2}\right) D\left( \phi /x\right) d^4x, 
\end{equation}
where $D(\phi /x)$ is the Jacobian defined as 
\begin{equation}
\in ^{ABCD}D\left( \phi /x\right) =\in ^{\mu \nu \lambda \rho }\partial _\mu
\phi ^A\partial \nu \phi ^B\partial _\lambda \phi ^C\partial _\rho \phi ^D. 
\end{equation}
By means of the general Green function formula 
\begin{equation}
\frac{\partial ^2}{\partial \phi ^A\partial \phi ^A}\left( \frac 1{||\phi
||^2}\right) =-4\pi \delta ^4\left( \vec \phi \right) , 
\end{equation}
we have 
\begin{equation}
C_2(P)=\delta ^4\left( \vec \phi \right) D\left( \phi /x\right) d^4x. 
\end{equation}

Suppose $\phi ^A(x)\ (A=0,1,2,3)$ have $m$ isolated zeros at $x_\mu =z_\mu
^i\ (i=1,2,\cdot \cdot \cdot ,m)$ , according to the $\delta -Function$
theory$^{28}$, $\delta (\vec \phi )$ can be expressed by 
\begin{equation}
\label{by}\delta (\vec \phi )=\sum_{i=1}^m\frac{\beta _i\delta (\vec x-\vec z%
_i)}{|D(\phi /x)|_{\vec x=\vec z_i}}, 
\end{equation}
and one then obtains 
\begin{equation}
C_2(P)=\sum_{i=1}^m\eta _i\beta _i\delta ^4\left( x-z_i\right) d^4x, 
\end{equation}
where $\beta _i$ is a positive integer (the Hopf index of the $i$th zeros )
and $\eta _i$ is the Brouwer degree$^{29}$: 
\begin{equation}
\eta _i=\frac{D\left( \phi /x\right) }{|D\left( \phi /x\right) |}%
=sgn[D\left( \phi /x\right) ]|_{x=z_i}=\pm 1. 
\end{equation}
The meaning of the Hopf index $\beta _i$ is that the vector field function $%
\vec \phi $ covers the corresponding region $\beta _i$ times while $\vec x$
covers the region neighborhood of zero $z_i$ once. From above discussion,
the Chern density $\rho (M)$ is defined as : 
\begin{equation}
\label{roi}\rho (M)=\sum_{i=1}^m\eta _i\beta _i\delta ^4\left( x-z_i\right)
, 
\end{equation}
which shows that the topological structure of Chern density $\rho $ is
labeled by the Brouwer degrees and the Hopf index. The integration of $\rho
(M)$ on $M$ 
\begin{equation}
\label{end}C_2=\int_M\rho (M)d^4x=\sum_{i=1}^m\eta _i\beta _i 
\end{equation}
is integer called Chern number which is a topological invariant of $M.$

The result (\ref{end}) suggest that the zeros points of the smooth vector $%
\vec \phi $ are essential to the topological properties of the base manifold 
$M.$ On the other hand, the density (\ref{roi}) can be regarded as the
density of a system of $k$ classical point-like particles with topological
invariant charges $g_i=\eta _i\beta _i$ on the 4-dimensional manifold.

\section*{IV. The Bifurcation of Chern Density}

As being discussed before, the zeros of the smooth vector $\vec \phi $ play
important roles in studying the Chern Class of the manifold $M$ . In this
section, we will study the properties of the zero points, in other words,
the properties of the following equations solutions 
\begin{equation}
\label{q1}\left\{ 
\begin{array}{c}
\phi ^0(x^0,x^1,x^2,x^3)=0 \\ 
\phi ^1(x^0,x^1,x^2,x^3)=0 \\ 
\phi ^2(x^0,x^1,x^2,x^3)=0 \\ 
\phi ^3(x^0,x^1,x^2,x^3)=0 
\end{array}
\right. . 
\end{equation}
As we knew before, if the Jacobian determinant%
$$
D(\frac \phi x)=\frac{\partial (\phi ^0,\phi ^1,\phi ^2,\phi ^3)}{\partial
(x^0,x^1,x^2,x^3)}\neq 0, 
$$
we will have the isolated solutions of (\ref{q1}). The isolated solution are
called regular points. It is easy to see that the result in section 3 is
based on this condition. However, when this condition fails, the above
results will change in some way, and will lead to the branch process of
topological density and give rise to the bifurcation.

In order to show this case easily, we want to have some suppose. It is
connivent to let $x^0=t$, and suppose $\phi ^0=$ $\phi ^0(t).$ We denote one
of zero points as ($t^{*},\vec z_i).$ We know that if 
$$
D(\frac{\vec \phi }{\vec x})|_{(t^{*},\vec z_i)}=\frac{\partial (\phi
^1,\phi ^2,\phi ^3)}{\partial (x^1,x^2,x^3)}|_{(t^{*},\vec z_i)}=0, 
$$
the Jacobian $D(\frac \phi x)|_{(t^{*},\vec z_i)}=0$ will be obtained
automatically. We will show that this case will lead to the branch process
of topological density. In this case, the equations(\ref{q1}) will be
rewritten as 
\begin{equation}
\label{q2}\phi ^0(t)=0, 
\end{equation}
and 
\begin{equation}
\label{q3}\left\{ 
\begin{array}{c}
\phi ^1(t,x^1,x^2,x^3)=0 \\ 
\phi ^2(t,x^1,x^2,x^3)=0 \\ 
\phi ^3(t,x^1,x^2,x^3)=0 
\end{array}
\right. . 
\end{equation}

It is well-known that when the Jacobian $D(\frac{\vec \phi }{\vec x}%
)|_{(t^{*},\vec z_i)}=0,$ the usual implicit function theorem is no use. But
if the Jacobian 
$$
D^1(\frac{\vec \phi }x)|_{(t^{*},\vec z_i)}=\frac{\partial (\phi ^1,\phi
^2,\phi ^3)}{\partial (t,x^2,x^3)}|_{(t^{*},\vec z_i)}\neq 0, 
$$
we can use the Jacobian $D^1(\frac{\vec \phi }x)|_{(t^{*},\vec z_i)}$
instead of $D(\frac{\vec \phi }{\vec x})|_{(t^{*},\vec z_i)}=0$ , for the
purpose of using the implicit function theorem$^{30}$. Then we have an
unique solution of the equations (\ref{q3}) in the neighborhood of the
points ($t^{*},\vec z_i$) 
$$
t=t(x^1) 
$$
\begin{equation}
\label{q4}x^i=x^i(x^1)\qquad i=2,3. 
\end{equation}
with $t^{*}=t(x^1).$ And we call the critical points ($t^{*},\vec z_i$) the
limit points. In the present case, it is easy to know that%
$$
\frac{dx^1}{dt}|_{(t^{*},\vec z_i)}=\frac{D^1(\frac{\vec \phi }x)|_{(t^{*},%
\vec z_i)}}{D(\frac{\vec \phi }x)|_{(t^{*},\vec z_i)}}=\infty 
$$
i.e.%
$$
\frac{dt}{dx^1}|_{(t^{*},\vec z_i)}=0. 
$$
Then we have the Taylor expansion of (\ref{q4}) at the point $\left( t^{*},%
\vec z_i\right) $%
$$
\begin{array}{c}
t=t^{*}+ 
\frac{dt}{dx^1}|_{(t^{*},\vec z_i)}(x^1-z_i^1)+\frac 12\frac{d^2t}{(dx^1)^2}%
|_{(t^{*},\vec z_i)}(x^1-z_i^1)^2 \\ =t^{*}+\frac 12\frac{d^2t}{(dx^1)^2}%
|_{(t^{*},\vec z_i)}(x^1-z_i^1)^2. 
\end{array}
$$
Therefore 
\begin{equation}
\label{q5}t-t^{*}=\frac 12\frac{d^2t}{(dx^1)^2}|_{(t^{*},\vec z%
_i)}(x^1-z_i^1)^2 
\end{equation}
which is a parabola in the $x^1-t$ plane. From (\ref{q5}), we can obtain the
two solutions $x_1^1(t)$ and $x_2^1(t),$ which give the branch solutions of
the system (\ref{q1}). If $\frac{d^2t}{(dx^1)^2}|_{(t^{*},\vec z_i)}>0,$ we
have the branch solutions for $t>t^{*},$ otherwise, we have the branch
solutions for $t>t^{*}.$ These two condition are related to the origin and
annihilation of topological charges$^{31}$. Since the Chern number (\ref{end}%
) is a topological invariant, the topological number of these two must be
opposite at the zero point, i.e.%
$$
\beta _{i1}\eta _{i1}=-\beta _{i2}\eta _{i2}. 
$$

For a limit point, it also requires the $D^1(\frac{\vec \phi }x)|_{(t^{*},%
\vec z_i)}\neq 0.$ As to a bifurcation point$^{32}$, it must satisfy a more
complement condition. This case will be discussed in the following
subsections in detail.

\subsection*{IV.1 The branch process at the bifurcation point.}

In this subsection, we have the restrictions of the system (\ref{q1}) at the
bifurcation point $(t^{*},\vec z_i)$%
\begin{equation}
\label{bifa12}\left\{ 
\begin{array}{c}
D( 
\frac{\vec \phi }{\vec x})|_{(t^{*},\vec z_i)}=0 \\ D^1(\frac{\vec \phi }x%
)|_{(t^{*},\vec z_i)}=0 
\end{array}
\right. 
\end{equation}

These will lead to an important fact that the function relationship between $%
t$ and $x^1$ is not unique in the neighborhood of the bifurcation point $(%
\vec z_i,t^{*})$. It is easy to see from the equation 
\begin{equation}
\label{bifa13}\frac{dx^1}{dt}\mid _{(t^{*},\vec z_i)}=\frac{D^1(\frac{\vec 
\phi }x)|_{(t^{*},\vec z_i)}}{D(\frac{\vec \phi }x)|_{(t^{*},\vec z_i)}} 
\end{equation}
which under the restraint (\ref{bifa12}) directly shows that the direction
of the integral curve of the equation (\ref{bifa13}) is indefinite at the
point $(\vec z_i,t^{*})$. This is why the very point $(\vec z_i,t^{*})$ is
called a bifurcation point of the system (\ref{q1}).

Next, we will find a simple way to search for the different directions of
all branch curves at the bifurcation point. Assume that the bifurcation
point $(\vec z_i,t^{*})$ has been found from (\ref{q3}) . We know that, at
the bifurcation point $(\vec z_i,t^{*})$, the rank of the Jacobian matrix $[%
\frac{\partial \vec \phi }{\partial x}]$ is smaller than $3$. First, we
suppose the rank of the Jacobian matrix $[\frac{\partial \vec \phi }{%
\partial x}]$ is $2$ (the case of a more smaller rank will be discussed
later). Suppose that the $2\times 2$ submatrix $J_1(\frac \phi x)\,$ is 
\begin{equation}
\label{bifa14}J_1(\frac \phi x)=\left( 
\begin{array}{cc}
\frac{\partial \phi ^1}{\partial x^2} & \frac{\partial \phi ^1}{\partial x^3}
\\ \frac{\partial \phi ^2}{\partial x^2} & \frac{\partial \phi ^2}{\partial
x^3} 
\end{array}
\right) , 
\end{equation}
and its determinant $D_1(\frac \phi x)$ does not vanish. The implicit
function theorem says that there exist one and only one function relation 
\begin{equation}
\label{bifa15}x^i=f^i(x^1,t),\quad i=2,3 
\end{equation}
We denoted the partial derivatives as%
$$
f_1^i=\frac{\partial f^i}{\partial x^1};\,\quad \,f_t^i=\frac{\partial f^i}{%
\partial t};\quad f_{11}^i=\frac{\partial ^2f^i}{\partial x^1\partial x^1}%
;\quad f_{1t}^i=\frac{\partial ^2f^i}{\partial x^1\partial x^t};\quad
f_{tt}^i=\frac{\partial f^i}{\partial x^t\partial x^t} 
$$
From (\ref{q1}) and (\ref{bifa15}) we have for $a=1,2,3$ 
\begin{equation}
\label{bifa16}\phi ^a=\phi ^a(x^1,f^2(x^1,t),f^3(x^1,t),t)=0 
\end{equation}
which give 
\begin{equation}
\label{bifa18}\frac{\partial \phi ^a}{\partial x^1}=\phi _1^a+\sum_{j=2}^3%
\frac{\partial \phi ^a}{\partial f^j}\frac{\partial f^j}{\partial x^1}=0 
\end{equation}
\begin{equation}
\label{bifa19}\frac{\partial \phi ^a}{\partial t}=\phi _t^a+\sum_{j=2}^3%
\frac{\partial \phi ^a}{\partial f^j}\frac{\partial f^j}{\partial t}=0. 
\end{equation}
from which we can get the first order derivatives of $f^i:\,f_1^i$ and $%
f_t^i $. Differentiating (\ref{bifa18}) with respect to $x^1$ and $t$
respectively we get 
\begin{equation}
\label{bifa21}\sum_{j=2}^3\phi _j^af_{11}^j=-\sum_{j=2}^3[2\phi
_{j1}^af_1^j+\sum_{k=2}^3(\phi _{jk}^af_1^k)f_1^j]-\phi _{11}^a\quad \quad
a=1,2,3 
\end{equation}
\begin{equation}
\label{bifa23}\sum_{j=2}^3\phi _j^af_{1t}^j=-\sum_{j=2}^3[\phi
_{jt}^af_1^j+\phi _{j1}^af_t^j+\sum_{k=2}^3(\phi _{jk}^af_t^k)f_1^j]-\phi
_{1t}^a\quad \quad a=1,2,3 
\end{equation}
And the differentiation of (\ref{bifa19}) with respect to $t$ gives 
\begin{equation}
\label{bifa24}\sum_{j=2}^3\phi _j^af_{tt}^j=-\sum_{j=2}^3[2\phi
_{jt}^af_t^j+\sum_{k=2}^3(\phi _{jk}^af_t^k)f_t^j]-\phi _{tt}^a\quad \quad
\quad a=1,2,3 
\end{equation}
where 
\begin{equation}
\label{bifa25}\phi _{jk}^a=\frac{\partial ^2\phi ^a}{\partial x^j\partial x^k%
},\quad \quad \phi _{jt}^a=\frac{\partial ^2\phi ^a}{\partial x^j\partial t}%
. 
\end{equation}
The differentiation of (\ref{bifa19}) with respect to $x^1$ gives the same
expression as (\ref{bifa23}). By making use of the Gaussian elimination
method to (\ref{bifa23}), (\ref{bifa23}) and (\ref{bifa24}) we can find the
second order derivatives $f_{11}^i,f_{1t}^i$ and $f_{tt}^i$. The above
discussion does no matter to the last component $\phi ^3(\vec x,t)$. In
order to find the different values of $dx^1/dt$ at the bifurcation point $(%
\vec z_i,t^{*})$, let us investigate the Taylor expansion of $\phi ^3(\vec x%
,t)$ in the neighborhood of $(\vec z_i,t^{*})$. Substituting (\ref{bifa15})
into $\phi ^3(\vec x,t)$ we have the function of two variables $x^1$ and $t$ 
\begin{equation}
\label{bifa26}F(x^1,t)=\phi ^3(x^1,f^2(x^1,t),f^3(x^1,t),t) 
\end{equation}
which according to (\ref{q3}) must vanish at the bifurcation point 
\begin{equation}
\label{bifa27}F(z_i^1,t^{*})=0. 
\end{equation}
From (\ref{bifa26}) we have the first order partial derivatives of $F(x^1,t)$
\begin{equation}
\label{bifa28}\frac{\partial F}{\partial x^1}=\phi _1^3+\sum_{j=2}^3\phi
_j^3f_1^j,\quad \quad \frac{\partial F}{\partial t}=\phi
_t^3+\sum_{j=2}^3\phi _j^3f_t^j. 
\end{equation}
Using (\ref{bifa18}) and (\ref{bifa19}) the first equation of (\ref{bifa12})
is expressed as 
\begin{equation}
\label{bifa29}D(\frac{\vec \phi }x)|_{(\vec z_i,t^{*})}=\left| 
\begin{array}{ccc}
-\sum\limits_{j=2}^3\phi _j^1f_1^j & \phi _2^1 & \phi _3^1 \\ 
-\sum\limits_{j=2}^3\phi _j^2f_1^j & \phi _2^2 & \phi _3^2 \\ 
\phi _1^3 & \phi _2^3 & \phi _3^3 
\end{array}
\right| _{(\vec z_i,t^{*})}=0 
\end{equation}
which by Cramer's rule can be written as 
$$
D(\frac{\vec \phi }x)|_{(\vec z_i,t^{*})}=\frac{\partial F}{\partial x^1}%
\det J_1(\frac \phi x)|_{((\vec z_i,t^{*})}=0 
$$
Since $detJ_1(\frac \phi x)|_{((\vec z_i,t^{*})}\neq 0$, the above equation
gives 
\begin{equation}
\label{bifa31}\frac{\partial F}{\partial x^1}\mid _{(\vec z_i,t^{*})}=0. 
\end{equation}
With the same reasons, we have 
\begin{equation}
\label{bifa32}\frac{\partial F}{\partial t}\mid _{(\vec z_i,t^{*})}=0. 
\end{equation}
The second order partial derivatives of the function $F$ are easily to find
out to be 
\begin{equation}
\label{bifa33}\frac{\partial ^2F}{(\partial x^1)^2}=\phi
_{11}^3+\sum\limits_{j=2}^3[2\phi _{1j}^3f_1^j+\phi
_j^3f_{11}^j+\sum\limits_{k=2}^3(\phi _{kj}^3f_1^k)f_1^j] 
\end{equation}
\begin{equation}
\label{bifa34}\frac{\partial ^2F}{\partial x^1\partial t}=\phi
_{1t}^3+\sum\limits_{j=2}^3[\phi _{1j}^3f_t^j+\phi _{tj}^3f_1^j+\phi
_j^3f_{1t}^j+\sum\limits_{k=2}^3(\phi _{jk}^3f_t^k)f_1^j] 
\end{equation}
\begin{equation}
\label{bifa35}\frac{\partial ^2F}{\partial t^2}=\phi
_{tt}^3+\sum\limits_{j=2}^3[2\phi _{jt}^3f_t^j+\phi
_j^3f_{tt}^j+\sum\limits_{k=2}^3(\phi _{jk}^3f_t^k)f_t^j] 
\end{equation}
which at $(\vec z_i,t^{*})$ are denoted by 
\begin{equation}
\label{bifa36}A=\frac{\partial ^2F}{(\partial x^1)^2}\mid _{(\vec z%
_i,t^{*})},\quad \quad B=\frac{\partial ^2F}{\partial x^1\partial t}\mid _{(%
\vec z_i,t^{*})},\ \quad \quad C=\frac{\partial ^2F}{\partial t^2}\mid _{(%
\vec z_i,t^{*})}. 
\end{equation}
Then taking notice of ( \ref{bifa27}), (\ref{bifa31}), (\ref{bifa32}) and (%
\ref{bifa36}) the Taylor expansion of $F(x^1,t)$ in the neighborhood of the
bifurcation point $(\vec z_i,t^{*})$ can be expressed as 
\begin{equation}
\label{bifa37}F(x^1,t)=\frac 12A(x^1-z_i^1)^2+B(x^1-z_i^1)(t-t^{*})+\frac 12%
C(t-t^{*})^2 
\end{equation}
which by (\ref{bifa26}) is the expression of $\phi ^3(\vec x,t)$ in the
neighborhood of $(\vec z_i,t^{*})$. The expression (\ref{bifa37}) is
reasonable, which shows that at the bifurcation point $(\vec z_i,t^{*})$ one
of the equations (\ref{q1}), $\phi ^3(\vec x,t)=0,$ is satisfied, i.e. 
\begin{equation}
\label{bifa38}A(x^1-z_i^1)^2+2B(x^1-z_i^1)(t-t^{*})+C(t-t^{*})^2=0. 
\end{equation}
Dividing (\ref{bifa38}) by $(t-t^{*})^2$ and taking the limit $t\rightarrow
t^{*}$ as well as $x^1\rightarrow z_i^1$ respectively we get 
\begin{equation}
\label{bifb38}A(\frac{dx^1}{dt})^2+2B\frac{dx^1}{dt}+C=0. 
\end{equation}
In the same way we have 
\begin{equation}
\label{bifa39}C(\frac{dt}{dx^1})^2+2B\frac{dt}{dx^1}+A=0. 
\end{equation}
The different directions of the branch curves at the bifurcation point are
determined by (\ref{bifb38}) or (\ref{bifa39}). It is easy to see that there
are at most two different branches.

\subsection*{IV.2 The branch process at a higher degenerated point.}

In the subsection $3.2$, we have studied the case that the rank of the
Jacobian matrix $[\frac{\partial \phi }{\partial x}]$ of the equations (\ref
{q1}) is $2=3-1$. In this subsection, we consider the case that the rank of
the Jacobian matrix is $1=3-2$. Let the $J_2(\frac{\vec \phi }x)=\frac{%
\partial \phi ^1}{\partial x^1}\,$ and suppose that $detJ_2\neq 0$. With the
same reasons of obtaining (\ref{bifa15}), we can have the function relations 
\begin{equation}
\label{bifa45}x^3=f^3(x^1,x^2,t)\quad . 
\end{equation}
Substituting the relations (\ref{bifa45}) into the last two equations of (%
\ref{q1}), we have the following two equations with three arguments $%
x^1,x^2,t$ 
\begin{equation}
\label{bifa46}\left\{ 
\begin{array}{l}
F_1(x^1,x^2,t)=\phi ^2(x^1,x^2,f^3(x^1,x^2,t),t)=0 \\ 
F_2(x^1,x^2,t)=\phi ^3(x^1,x^2,f^3(x^1,x^2,t),t)=0. 
\end{array}
\right. 
\end{equation}
Calculating the partial derivatives of the function $F_1$ and $F_2$ with
respect to $x^1$, $x^2$ and $t$, taking notice of (\ref{bifa45}) and using
six similar expressions to (\ref{bifa31}) and (\ref{bifa32}), i.e. 
\begin{equation}
\label{bifa48}\frac{\partial F_j}{\partial x^1}\mid _{(\vec z_i,t^{*})}=0,\
\quad \quad \frac{\partial F_j}{\partial x^2}\mid _{(\vec z_i,t^{*})}=0,\
\quad \quad \frac{\partial F_j}{\partial t}\mid _{(\vec z_i,t^{*})}=0,\
\quad \quad j=1,2, 
\end{equation}
we have the following forms of Taylor expressions of $F_1$ and $F_2$ in the
neighborhood of $(\vec z_i,t^{*})$ 
$$
F_j(x^1,x^2,t)\approx
A_{j1}(x^1-z_i^1)^2+A_{j2}(x^1-z_i^1)(x^2-z_i^2)+A_{j3}(x^1-z_i^1) 
$$
$$
(t-t^{*})+A_{j4}(x^2-z_i^2)^2+A_{j5}(x^2-z_i^2)(t-t^{*})+A_{j6}(t-t^{*})^2=0 
$$
\begin{equation}
\label{bifa49}j=1,2. 
\end{equation}
In case of $A_{j1}\neq 0,A_{j4}\neq 0$, dividing (\ref{bifa49}) by $%
(t-t^{*})^2$ and taking the limit $t\rightarrow t^{*}$, we obtain two
quadratic equations of $\frac{dx^1}{dt}$ and $\frac{dx^2}{dt}$ 
\begin{equation}
\label{bifa50}A_{j1}(\frac{dx^1}{dt})^2+A_{j2}\frac{dx^1}{dt}\frac{dx^2}{dt}%
+A_{j3}\frac{dx^1}{dt}+A_{j4}(\frac{dx^2}{dt})^2+A_{j5}\frac{dx^2}{dt}%
+A_{j6}=0 
\end{equation}
$$
j=1,2. 
$$
Eliminating the variable $dx^1/dt$, we obtain a equation of $dx^2/dt$ in the
form of a determinant 
\begin{equation}
\label{bifa51}\left| 
\begin{array}{cccc}
A_{11} & A_{12}v+A_{23} & A_{14}v^2+A_{15}v+A_{16} & 0 \\ 
0 & A_{11} & A_{12}v+A_{13} & A_{14}v^2+A_{15}v+A_{16} \\ 
A_{21} & A_{22}v+A_{23} & A_{24}v^2+A_{25}v+A_{26} & 0 \\ 
0 & A_{21} & A_{22}v+A_{23} & A_{24}v^2+A_{25}v+A_{26} 
\end{array}
\right| =0 
\end{equation}
where $v=dx^2/dt$, that is a $4th$ order equation of $dx^2/dt$ 
\begin{equation}
\label{bifa52}a_0(\frac{dx^2}{dt})^4+a_1(\frac{dx^2}{dt})^3+a_2(\frac{dx^2}{%
dt})^2+a_3(\frac{dx^2}{dt})+a_4=0. 
\end{equation}
Therefore we get different directions at the bifurcation point corresponding
to different branch curves. The number of different branch curves is at most
four. At the end of this section, we conclude that in our theory of
topological density there exist the crucial case of branch process. This
means that a normal point-like topological charge, when moves through the
bifurcation point, may split into several point-like topological charges
moving along different branch curves. Since the topological density is a
invariant, the total charge of the spliting topological particles must
precisely equal to the topological charge of the original particle.

Since the Chern density is identically conserved, the sum of the topological
charges of these splitted topological density must be equal to that of the
original current at the bifurcation point. We suppose that there exist $l$
different branches. Then the topological density of Chern class $\rho (M)$
changes in the following form%
$$
\rho (M)=\sum_{i=1}^k\rho _i=\sum_{i=1}^k\sum_{j=1}^l\rho _{ij} 
$$
where 
\begin{equation}
\label{rob}
\begin{array}{c}
\rho _i=g_i\delta ^4\left( x-z_i\right)
,\,\,\,\,\,\,\,\,\,\,\,\,\,\,\,\,\,\,\,\,\,\,\rho _{ij}=g_{ij}\delta
^4\left( x-z_{ij}\right) , \\ 
1\leq j\leq l. 
\end{array}
\end{equation}
With the same reason, the sum of the topological charges at the bifurcation
points will be have the following form 
$$
g_i=\sum_{i=1}^lg_{ij}. 
$$
for fixed $i$.

\section*{V. Conclusion}

We have explicitly constructed the gauge potential decomposition theory of $%
SU(2)$ gauge theory in terms of the sphere bundle on a 4-dimensional
manifold. An important observation of the application of this theory is that
the $SU(2)$ Chern density takes the form of a general function. And we find
that the Chern density have the bifurcation process. It is shown that the
topological charges are splitted under this case, and the total charges is
conserved. Those features are very important in discussing the topological
problems on Four-Manifold. Since many problems, not only in theoretical
physics, but also in differential geometry, are associated with the $SU(2)$
gauge theory, the theory of $SU(2)$ gauge potential decomposition, the inner
structure of the $SU(2)$ Chern density will provide an important and
powerful methods in those fields.

\section*{Acknowledgments}

This work was supported by the National Natural Science Foundation of China.

$^1$Y.S. Duan and S.L{\bf . }Zhang{\bf ,} Int. J. Eng. Sci.\ {\bf 28 }689
(1990){\bf ; 29 }153 (1991); Int. J. Eng. Sci.{\bf \ 30\ }153 (1992).

$^2$Y.S. Duan, S.L. Zhang and S.H{\bf . }Feng{\bf , }J. Math.Phys. {\bf 35 }%
9 (1994).

$^3$Y.S. Duan and X.H. Meng,\ J. Math. Phys.\ {\bf 34} 1149 (1993); Y.S.
Duan and Lee, X.G{\bf .:} Helv. Phys. Acta.{\bf \ 68 }513 (1995).

$^4$Y.S. Duan, G.H. Yang and Y. Jing, Helv.Phys.Acta.(to be published).

$^5$J.S. Dowker and J.P{\bf . }Schofield{\bf ,} J. Math. Phys. {\bf 31 }808
(1990).

$^6$Y.S. Duan and M.L{\bf . }Ge{\bf ,} Kexue Tongbao{\bf \ 26 }282 (1976);
Sci. Sinica{\bf \ 11} 1072 (1979); Y.S. Duan and S.C. Zhao,\ Commun. Theory
Phys.{\bf \ 2 }1553 (1983).{\bf \ }

$^7$Y.S. Duan,{\bf \ }SLAC-PUB-3301/84.

$^8$Y.S. Duan and J.C. Liu, Proceedings of Johns Hopkins Workshop 11, edited
by Y.S. Duan, G. Domokos and Kovesi-Domokos, S. (World Scientific,
Singapore, 1988).

$^9$J.M.F.Labastida and M.Marino, Nucl.Phys. {\bf B456} 633 (1996).

$^{10}$J.Wess and B.Zumino, Phys.Lett. {\bf B37 } 95 (1971).

$^{11}$E.Witten, {\it Monopoles and Four-Manifold},
hep-th/9411102,IASSNS-HEP-94-96.

$^{12}$E.Witten, J.Math.Phys. {\bf 35 }5101 (1994).

$^{13}$C.Michael and P.S.Spencer, Phys.Rev.D. {\bf 52 } 4691 (1996).

$^{14}$A.Galperin and E.Sokatchev, Report BONN-TH-94-27 24 (1994).

$^{15}$P.M.Sotcliffe, Nucl.Phys.{\bf \ B431 }97 (1995).

$^{16}$Alfred Actor, Rev.Mod.Phys. {\bf 51 }461 (1979).

$^{17}$V.Alfaro.et.al,Nucl.Phys. {\bf B73 }463 (1978).

$^{18}$S.Donaldson, Topology, {\bf 29 }257 (1990).

$^{19}$D. Hestenes and G. Sobczyk,\ {\it Clifford Algebra to Geometric
Calculus} (Reidel, Dordrecht 1984).

$^{20}$S.Okubo, J.Math.Pys.{\bf 32 }1657 (1991).

$^{21}$S.S. Chern,\ An. da Acad. Brasileira de Ci\'encias, {\bf 35 }17\
(1963).

$^{22}$S. Nash and S{\bf . }Sen{\bf ,} {\it Topological and Geometry of
Physicists}{\bf \ (}Academic Pre. INC. London. 1983); M.W{\bf . }Hirsch{\bf %
, }{\it Differential Topology} (Springer Verlag. New York 1976).

$^{23}$C.Doran,D.Hestenes,F.Sommen and N.V.Acker, J.Math.Phys. {\bf 34 }3642
(1993);

A.Lasenby,C.Doran and S.Gull, J.Math.Phys. {\bf 34 }3683 (1993).

$^{24}$H.Boerner, {\it Representation of Groups} (North-Holland Publishing
Company 1963).

$^{25}$A. Dimakis and M\"uller-Hoisen, Class.Quantum Grav. {\bf 8 }2093
(1991).

$^{26}$Eguchi,Gilkey and Hanson, {\it Gravitation, gauge theories and
differential geometry}, Phys.Rep. {\bf 66} 213 (1980).

$^{27}$S.S. Chern and J. Simons, Ann.Math. {\bf 99 }48 (1974).

$^{28}$Albert S.Schwarz, {\it Topology For Physicist} (Springer Verlag Press
1994).

$^{29}$H. Hopf,\ Math. Ann. {\bf 96} 209 (1929).

$^{30}$E.Goursat, {\it A Course in Mathematical Analysis} {\bf Vol. 1}%
(transl. E. R. Hedrick).

$^{31}$Nobuyuki Sakai, Phys. Rev. {\bf D54} 1548 (1996).

$^{32}$M. Kubicek, and M. Marek, {\it Computational Methods in Bifurcation
Theory and Dissipative Structures} (Springer-Verlag, New York 1983).

\end{document}